\documentclass[fleqn,twoside,twocolumn,nofootinbib,showkeys]{revtex4}
\usepackage[sec,nopacs,nocpr]{ujp}
\usepackage{nccmath}
\usepackage{placeins}
\usepackage{hyperref} %the build LaTeX => PDF [pdftex,colorlinks=true,allcolors=blue,breaklinks=true]

\begin{document}
	\title[Magnon Bose-Einstein condensate and supercurrents over a wide temperature range]{MAGNON BOSE-EINSTEIN CONDENSATE AND SUPERCURRENTS OVER A WIDE TEMPERATURE RANGE}
	\author{L.~Mihalceanu}
	\affiliation{Fachbereich Physik and Landesforschungszentrum OPTIMAS, Technische Universit\"at Kaiserslautern}
	\address{67663 Kaiserslautern, Germany}
	\email{mihalcea@rhrk.uni-kl.de}
	\author{D.A.~Bozhko}
	\affiliation{Fachbereich Physik and Landesforschungszentrum OPTIMAS, Technische Universit\"at Kaiserslautern}
	\address{67663 Kaiserslautern, Germany}
	\affiliation{James Watt School of Engineering, University of Glasgow}
	\address{Glasgow G12 8LT, United Kingdom}
	%\email{dmytro.bozhko@glasgow.ac.uk}
	%\affiliation{Department of Physics and Energy Science, University of Colorado at Colorado Springs}
	%\address{Colorado Springs CO 80918, USA}
	\email{dbozhko@uccs.edu}
	\author{V.I.~Vasyuchka}
	\affiliation{Fachbereich Physik and Landesforschungszentrum OPTIMAS, Technische Universit\"at Kaiserslautern}
	\address{67663 Kaiserslautern, Germany}
	\email{vasyuchka@physik.uni-kl.de}
	\author{A.A.~Serga}
	\affiliation{Fachbereich Physik and Landesforschungszentrum OPTIMAS, Technische Universit\"at Kaiserslautern}
	\address{67663 Kaiserslautern, Germany}
	\email{serga@physik.uni-kl.de}
	\author{B.~Hillebrands}
	\affiliation{Fachbereich Physik and Landesforschungszentrum OPTIMAS, Technische Universit\"at Kaiserslautern}
	\address{67663 Kaiserslautern, Germany}
	\email{hilleb@physik.uni-kl.de}
	\author{A.~Pomyalov}
	\affiliation{Department of Chemical and Biological Physics, Weizmann Institute of Science }
	\address{Rehovot 76100, Israel}
	\email{Anna.Pomyalov@weizmann.ac.il}
	
	\author{V.S.~L'vov}
	\affiliation{Department of Chemical and Biological Physics, Weizmann Institute of Science }
	\address{Rehovot 76100, Israel}
	\email{victor.lvov@gmail.com}
	\author{V.S.~Tiberkevich}
	\affiliation{Department of Physics, Oakland University }
	\address{Rochester MI 48309, United States}
	\email{tyberkev@oakland.edu}
	
	%\udk{539} \pacs{75.30.Ds, 75.45.+j,\\ 03.75.Kk, 75.50.Gg}
	\razd{\secviii}

	\autorcol{L.~Mihalceanu, D.A.~Bozhko, V.I.~Vasyuchka et al.}
	
	\setcounter{page}{1}%

	\begin{abstract}
		Magnon Bose-Einstein Condensates (BECs) and supercurrents are coherent quantum phenomena, which appear on a macroscopic scale in parametrically populated solid state spin systems. One of the most fascinating and attractive features of these processes is the possibility of magnon condensation and supercurrent excitation even at room temperature. At the same time, valuable information about a magnon BEC state, such as its lifetime, its formation threshold, and coherency, is provided by experiments at various temperatures. Here, we use Brillouin Light Scattering (BLS) spectroscopy for the investigation of the magnon BEC dynamics in a single-crystal film of yttrium iron garnet in a wide temperature range from 30\,K to 380\,K. By comparing the BLS results with previous microwave measurements, we revealed the direct relation between the damping of the condensed and the parametrically injected magnons. The enhanced supercurrent dynamics was detected at 180\,K near the minimum of BEC damping.
	\end{abstract}
	
	\keywords{magnon gas, parametric pumping, Bose-Einstein condensate, magnon superfluidity, magnon supercurrent, yttrium iron garnet (YIG).}
	
	\maketitle
	
	\section{Introduction}
	The phenomena related to room-temperature Bose-Einstein condensation \cite{Melkov1994,Demokritov2006,Bugrij2007,Rezende2009,Serga2014,Bunkov2018} and superfluidity \cite{Nowik-Boltyk2012,Bozhko2016,Bozhko2019_1} in overpopulated magnon gases constitute a fast evolving field of research, revealing promising features for advanced technological applications \cite{Chumak2015} and discovery of novel physical effects \cite{Nakata2014,Nakata2015,Skarsvag2015,Sugakov2016,Flebus2016,Tiberkevich2019}. Recent examples of such phenomena are magnon supercurrents \cite{Bozhko2016}, Bogoliubov \cite{Bozhko2019_1} and second sound \cite{Tiberkevich2019} waves, microscaled vorticity \cite{Nowik-Boltyk2012} in a magnon condensate, thermally induced magnon Bose-Einstein Condensates (BECs) in microsized magnetic structures \cite{Safranski2017,Schneider2018}, and the interaction of the magnon BEC with  accumulated hybrid magnetoelastic bosons \cite{Bozhko2017}. The intrinsic coherency of magnon BECs, allowing for a phase-encoded information processing and a long-range supercurrent-carried data transfer in the GHz regime, creates the potential for applications of these macroscopic quantum phenomena in the field of wave-based computing.
	
	Here, we investigate the process of Bose-Einstein condensation of parametrically pumped magnons in a single-crystal Yttrium Iron Garnet \cite{Cherepanov1993} (YIG, $\mathrm{Y_{3}Fe_{5}O_{12}}$) film in a wide range of temperatures (from 380\,K down to 30\,K). The temperature dependence of the  magnon BEC decay rate is in quantitative agreement with the temperature dependence of the relaxation rates of the  parametrically injected magnons \cite{Mihalceanu2018}. In particular, both relaxation rates exhibit a non-monotonous behavior, with a minimum at around 200\, K and a large increase for temperatures below 100\,K. The magnon supercurrents, which in our experiments manifest themselves in the enhanced magnon BEC decay, were found to be most prominent around 180\,K, i.e., in the regime where the magnon damping is minimal.
	
	\section{Experimental Setup}
	
\begin{figure}[b]% figure* for wide figure, [h] [!] to change the placement
		\includegraphics[width=\columnwidth]{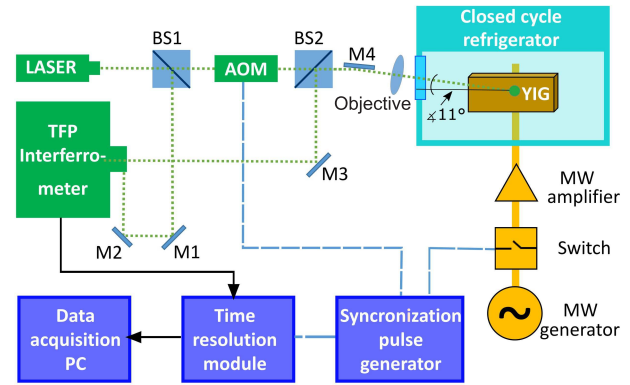}
		\caption{Sketch of the experimental setup composed of a Brillouin light scattering spectroscopy part (green), microwave excitation circuit (yellow), temperature control part (cyan), and time synchronization and data acquisition part (blue). M1-M4 -- mirrors. BS1 -- beam splitter. BS2 -- polarizing beam splitter. }
		\label{fig1}
	\end{figure}

	To investigate the temperature dependence of  the magnon condensate properties, we designed a setup composed of a microwave circuit for the parametric excitation of magnons and a Brillouin Light Scattering (BLS) spectroscopy part for magnon detection \cite{Serga2012} (see Fig.\,\ref{fig1}). The YIG sample is placed on top of a microstrip resonator, attached from below to a highly heat-conducting $\mathrm{AlN}$ substrate. This part of the setup is mounted inside of a Janis Research 10\,K closed cycle refrigerator system with optical access.
	
	%Fig.~1
	
	For the parametric pumping process, a microwave (MW) pumping pulse of 2\,$\mu$s duration at a frequency $\omega_{\mathrm{p}} = 2\pi \cdot 15$\,{GHz} with an 1\,ms repetition rate and a maximal pumping power $P_{\mathrm{p}}$ of 12\,{W} was applied to a microwave transmission line capacitively coupled to a 50\,$\mu$m-wide microstrip resonator. The microwave Oersted field ${\bf h_{\mathrm{p}}}$ induced by the microstrip resonator
	parametrically drives the magnetization of the YIG-film sample \cite{Neumann2009}. The parametric pumping process \cite{Gurevich_Melkov_book,Melkov1999} can be understood as a decay of a microwave photon with frequency $\omega_{\mathrm{p}}$ and nearly zero wavevector into two magnons at half the photon frequency $\omega_{\mathrm{p}}/2 = 2 \pi \cdot 7.5$\,{GHz} having opposite wavevectors $\mathbf{q}$ and $-\mathbf{q}$.
	
	After thermalization \cite{Demidov2007,Hick2012,Clausen2015} of the parametrically injected magnons over the spin-wave spectrum, a quasi-equilibrium distribution of the magnon gas is established. The effective chemical potential of such a quasiparticle system is nonzero and depends on the density of the injected excessive magnons. At a sufficiently large pumping power the chemical potential reaches the bottom of the magnon spectrum and Bose-Einstein condensation of magnons takes place \cite{Demokritov2006,Serga2014,Demidov2008,Bozhko2015}.
	
	The time-resolved BLS spectroscopy measurements were performed in a backward scattering geometry \cite{Demokritov2001}. In such a geometry, the incidence angle of the probing laser beam can be adjusted to selectively detect magnons with a particular value of the in-plane wavevector  \cite{Sandweg2010}. In the experimental data shown below, the incidence angle was fixed at $11^\circ$, which corresponds to the detection of magnons at the bottom of the spin-wave spectrum, where thermalized magnons accumulate and the magnon BEC is formed.
	
\begin{figure}[t]% figure* for wide figure, [h] [!] to change the placement
		\includegraphics[width=\columnwidth]{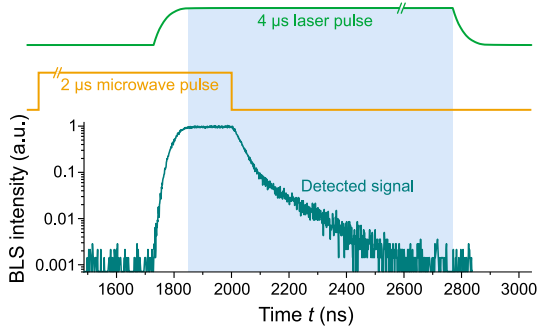}   
		\caption{Schematics of the relevant time intervals. Top: the 2$\mathrm{\mu s}$ long MW pumping pulse (yellow) and the 4$\mathrm{\mu s}$ long pulsed BLS laser beam (green). Bottom: the plot of the resulting detected BLS photon counts (magnon density) over time, detected at 180 K. The shaded area shows the part of the recorded data traces used for the analysis.}
		\label{fig2}
	\end{figure}

	In order to reduce the optical heating of the YIG sample, the probing laser beam was chopped by an Acousto-Optic Modulator (AOM) into a sequence of light pulses of 4\,$\mu$s duration (see Fig.\,\ref{fig1}). In the time-resolved experiments, the probing laser pulse has been applied 1.8\,$\mu$s after the start of the microwave pumping pulse. This scheme allowed us to map time evolution of the magnon BEC during both the pumping and free-decay stages (see Fig.\,\ref{fig2}).
	
	%Fig.~2
	
	%Fig.~3
	\begin{figure}[t]% figure* for wide figure, [h] [!] to change the placement
		\includegraphics[width=\columnwidth]{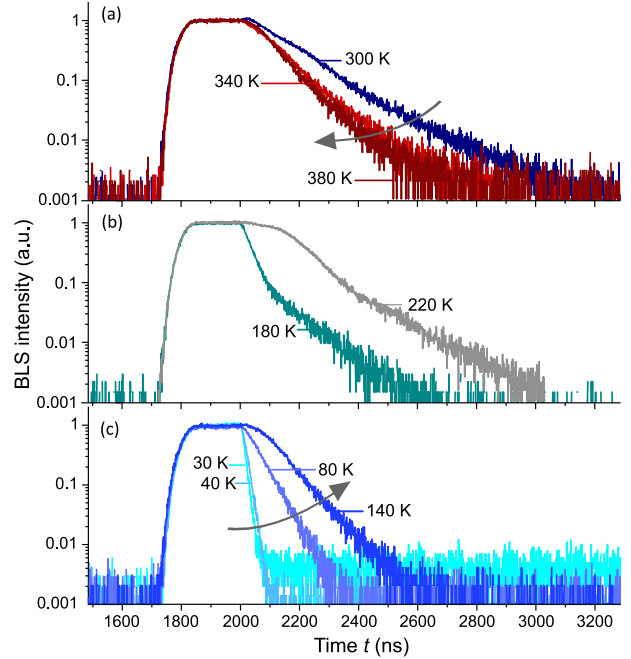}
		\caption{Time profiles of the BLS intensity for different temperatures: (a) 380\,K to 300\,K, (b) 220\,K and 180\,K, and (c) 140\,K to 30\,K. In panels (a) and (c), the temperature increases in the direction of the gray arrows. The decay characteristics of the profiles at $t > 2100$\,ns provide information about the dissipation and redistribution of the magnon BEC.}
		\label{fig3}
	\end{figure}
	
	The YIG sample used in the experiments has lateral dimensions of $1\times5\,\mathrm{mm^2}$ and a thickness of 5.6~$\mathrm{\mu m}$. It was prepared by chemical etching of an YIG film epitaxially grown in the (111) crystallographic plane on a GGG substrate of $500\,\mathrm{\mu m}$ thickness. The sample was magnetized along its long axis to avoid undesirable influence of static demagnetizing field on the value of the internal magnetic field.
	
	To achieve the most efficient condensation of magnons, we pumped the magnon gas at an applied magnetic field $H$ corresponding to the kinetic instability regime \cite{Lavrinenko1981,Melkov1991,Kreil2018}.
	% (more processes involved, kolmogorov cascade)
	Since the YIG saturation magnetization $M_\mathrm{S}$ is temperature-dependent, the external field was adjusted at each temperature to maximize the BLS signal {\cite{Mihalceanu2018}. Thus, the measurements have been performed at fields ranging from 1140\,Oe to 1250\,Oe.
		
		The temporal profiles of the BEC intensity, measured in the temperature range  30--380\,K, are shown in Fig.\,\ref{fig3}. To facilitate a comparison of the magnon decay processes at different temperatures, the BEC profiles have been normalized to their maximum intensity measured during the action of the microwave pumping pulse (the plateau region).
		In this work, we are mostly interested in the free decay of magnon BEC, i.e. in the part of the temporal profiles measured after the pumping pulse was switched off ($t = 2000$\,ns in Fig.~\ref{fig3}). Below we show that  the time dependencies of the magnon intensity decay  provide valuable information about the dissipation and the coherence properties of the magnon BEC.
		%
		%
		%the magnons decay to YIG lattice and redistribute over the magnon  with the rate, which can be found from the slope of the experimental curves. is revealed by the slope of the time pulse. After the states have been populated by the pumped magnons and the pumping process stops, the BLS photons scatter on the remaining magnons, as these redistribute and decay. The characteristics of this redistribution allows to understand the underlying process and obtain information about the various mechanisms.
		
		\section{Results}
		The temporal evolutions of the BLS intensities, measured at different temperatures, are shown in  Fig.\,\ref{fig3}.
		
		At relatively high temperatures $T \gtrsim 200$\,K (see Fig.\,\ref{fig3}a, b), the rate of the magnon BEC decay increases with rising temperature. At low temperatures $T \lesssim 200$\,K, however, the trend changes and the magnon BEC decay rate starts to increase as the temperature goes down (see Fig.\,\ref{fig3}c). Thus, the decay rate of the magnon BEC exhibits a clear minimum at temperatures around $T \approx 200$\,K. The increase of the decay rate at higher temperatures $T > 200$\,K can be explained by the a scattering of the condensed coherent magnons on the  higher-energy thermally-excited incoherent magnons. The increase of the damping at low temperatures is, most probably, related to the magnon absorption in the paramagnetic GGG substrate and to their scattering at magnetic impurities in a YIG film \cite{Mihalceanu2018,Danilov1989,Boventer2018,Kosen2019}.
		
		An interesting property of the time evolution of the of magnon BEC intensity is the observed two-stage decay of the BEC density. It is most clearly visible for $T = 180$\,K (see Fig.\,\ref{fig3}b), where the magnon decay rate is near its minimal value, and becomes much less pronounced with the increase of magnon damping. The initial, non-exponential decay is not related to any dissipation mechanism, but is a signature of magnon supercurrents \cite{Bozhko2016,Kreil2018,Bozhko2019_1}, which lead to a spatial redistribution of the coherent magnon BEC \cite{Bozhko2019_1}. This coherent redistribution process is driven by a thermal gradient within the BEC phase, locally induced by the probing laser light.
		
		In the following analysis, we neglect the initial part of the decaying magnon BEC intensity profile $I(t)$, in which magnon supercurrents are important, and fit the remaining part with the exponential function
		%\begin{ceqn}
		%    \begin{align}
		\begin{equation}\label{fit}
		\hskip 3 cm    I(t) \sim e^{-2\Gamma t}\, ,
		\end{equation}
		%    \end{align}
		%\end{ceqn}
		where the fitting parameter $\Gamma$ has the meaning of a magnon BEC relaxation rate (i.e., relaxation rate of magnons at the bottom of the magnon spectrum) \cite{Gurevich_Melkov_book}. The results of this analysis are shown in Fig.\,\ref{fig4}, together with the temperature dependence of the relaxation rate of parametrically injected magnons, measured in Ref.\,\cite{Mihalceanu2018}. Here, the dimensionless data from Ref.\,\cite{Mihalceanu2018} were scaled to coincide with the BEC relaxation rate at one temperature point $T = 340$\,K.

		As is clearly seen from Fig.\,\ref{fig4}, the relaxation rates of both groups of magnons have the same qualitative and quantitative temperature dependence. The  magnon damping is minimal around $T \approx 200$\,K, slowly increases for higher temperatures, and demonstrates a large increase for $T < 150$\,K. At the same time, this low-temperature relaxation increase is visibly slower for the magnon BEC than for the parametric magnons excited in the same YIG film (solid yellow curve in Fig.\,\ref{fig4}). The behavior of the BEC relaxation rate is much closer to the  temperature dependence of parametric magnons measured in the 10 times thicker YIG film of $53\,\mu$m thickness (dashed orange curve in Fig.\,\ref{fig4}). This fact can be understood as the manifestation of a slightly weaker coupling of the condensed short-wavelength dipole-exchange magnons (BEC wavenumbers $q_\mathrm{BEC} \approx \pm 4\cdot 10^4\,\mathrm{rad}\,\mathrm{cm}^{-1}$) \cite{Serga2014} to the paramagnetic spin structure of the GGG substrate in comparison with such a coupling of the rather long-wavelength dipolar magnons ($q \rightarrow 0$) parametrically excited and characterized in work~\cite{Mihalceanu2018}. This coupling is expected to be smaller  in a thicker film than in a thinner one, and, thus, the BEC relaxation rates in the thinner film come closer to the relaxation of long-wavelength parametric magnons in the thicker one.

		%Fig.~4
		\begin{figure}[t]% figure* for wide figure, [h] [!] to change the placement
			\includegraphics[width=\column]{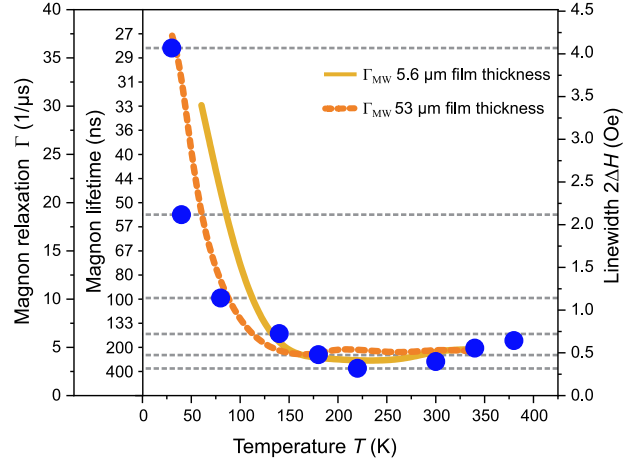}
			\caption{Temperature dependence of the magnon relaxation parameter $\Gamma$, lifetime $\tau=1/\Gamma$ (left axes), and the damping related linewidth $2 \Delta H = 2\Gamma/\gamma$ (right axis). Here $\gamma = 2\pi\cdot2.8$\,MHz/Oe is the modulus of the gyromagnetic ratio. Points -- damping of magnon BEC, measured in this work; lines -- damping of parametrically injected magnons in YIG films of two different thicknesses (from Ref.\,\cite{Mihalceanu2018}). }
			\label{fig4}
		\end{figure}
		
		\section{Conclusions}
		We have experimentally measured the temperature dependence of the relaxation rate of magnons at the bottom of the magnon spectrum (at the point of magnon Bose-Einstein condensation). We have found, that the temperature dependence of the damping of magnons in this spectral range is in quantitative agreement with the damping of another important group of short-wavelength magnons -- those, which are directly injected by microwave parametric pumping. Namely, the magnon damping have a minimum at $T \approx 200$\,K and increases for both higher and lower temperatures. We have also found that magnon BEC supercurrents are strongly enhanced in the temperature range near the minimum of magnon damping. The similar temperature behavior of the magnon BEC and parametrically injected magnons brings hope that an extremely long BEC lifetime can be achieved in ultra pure bulk YIG samples, where, as it has been found in Refs.\,\cite{Mihalceanu2018,Danilov1996}, the relaxation rate of parametric magnons monotonically decreases from room to cryogenic temperatures.

		\vskip3mm \textit{Financial support by the European Research Council within the Advanced Grant No. 694709 ``SuperMagnonics'', by Deutsche Forschungsgemeinschaft (DFG) within the Transregional Collaborative Research Center SFB/TR 49 ``Condensed Matter Systems with Variable Many-Body Interactions'', by the DFG Project No. INST 248/178-1, and by the NSF of the USA grants Nos. EFMA-1641989 and ECCS-1708982 is gratefully acknowledged. D.A.~Bozhko acknowledges support from the Alexander von Humboldt Foundation. The authors are grateful to G.A.~Melkov for fruitful discussions.}

		%%\vspace*{-5mm}
%		\rezume
%		{Ë. Ì³õàëü÷åàíó, Ä.À. Áîæêî, Â.². Âàñþ÷êà, Î.Î. Ñåðãà, Á. Õ³ëëåáðàíäñ, À. Ïîìÿëîâ, Â.Ñ. Ëüâîâ, Â.Ñ. Òèáåðêåâè÷}
%		%
%		{ÌÀÃÍÎÍÍÈÉ ÁÎÇÅ-ÅÉÍØÒÅÉÍIÂÑÜÊÈÉ \\ ÊÎÍÄÅÍÑÀÒ ÒÀ ÑÓÏÅÐÑÒÐÓÌÈ \\ Â ØÈÐÎÊÎÌÓ ÒÅÌÏÅÐÀÒÓÐÍÎÌÓ Ä²ÀÏÀÇÎÍ²}
%		%
%		{Ìàãíîíí³ Áîçå-Åéíøòåéí³âñüê³ êîíäåíñàòè òà ñóïåðñòðóìè º êîãåðåíòíèìè êâàíòîâèìè ÿâèùàìè, ùî ïðîÿâëÿþòüñÿ íà ìàêðîñêîï³÷íèõ ðîçì³ðàõ â ïàðàìåòðè÷íî çàñåëåíèõ ñï³íîâèõ ñèñòåìàõ òâåðäèõ ò³ë. Ìîæëèâ³ñòü ìàãíîííî¿ êîíäåíñàö³¿ òà çáóäæåííÿ ñóïåðñòðóì³â íàâ³òü çà ê³ìíàòíèõ òåìïåðàòóð º îäí³ºþ ç íàéá³ëüø çàõîïëþþ÷èõ òà ïðèâàáëèâèõ ðèñ öèõ ïðîöåñ³â. Âîäíî÷àñ, òåìïåðàòóðí³ äîñë³äæåííÿ ñïðîìîæí³ íàäàòè ö³ííó ³íôîðìàö³þ ïðî ìàãíîíí³ êîíäåíñàòè, çîêðåìà ïðî ¿õí³ ÷àñ æèòòÿ, ïîð³ã ôîðìóâàííÿ, òà ñòóï³íü êîãåðåíòíîñò³. Â ö³é ðîáîò³, ìè çàñòîñóâàëè ñïåêòðîñêîï³þ áð³ëëþåí³âñüêîãî ðîçñ³ÿííÿ ñâ³òëà äëÿ äîñë³äæåííÿ äèíàì³êè ìàãíîííîãî êîíäåíñàòó â ìîíîêðèñòàë³÷í³é ïë³âö³ çàë³çî-³òð³ºâîãî ãðàíàòó â øèðîêîìó ä³àïàçàí³ òåìïåðàòóð â³ä 30\,K äî 380\,K. Ïîð³âíþþ÷è ðåçóëüòàòè áð³ëëþåí³âñüêî¿ ñïåêòðîñêîï³¿ ç ïîïåðåäí³ìè íàäâèñîêî÷àñòîòíèìè âèì³ðàìè, ìè âèÿâèëè ïðÿìèé çâ'ÿçîê ì³æ çàòóõàííÿì çêîíäåíñîâàíèõ òà ïàðàìåòðè÷íî ³íæåêòîâàíèõ ìàãíîí³â. Ïðè òåìïåðàòóð³ 180\,K, ïîáëèçó ì³í³ìàëüíèõ çíà÷åíü çàòóõàííÿ Áîçå-Åéíøòåéí³âñüêîãî êîíäåíñàòó, áóëî çàðåºñòðîâàíî çíà÷íî ³íòåíñèô³êîâàíó äèíàì³êó ìàãíîííèõ ñóïåðñòðóì³â.}
	\end{document}